\newcommand{\ie}{{\it i.e.}}
\newcommand{\eg}{{\it e.g.}}

\newcommand{\rhs}{r.h.s.\ }

\newcommand{\BesselJ}{{\mathop{\rm J}\nolimits}}

\documentstyle[prb,aps]{revtex}

\draft

\begin{document}

\title{
  Cherenkov radiation from fluxon in a stack of coupled long Josephson junctions.
}

\author{
  E.~Goldobin\cite{email}
}

\address{
  Institute of Thin Film and Ion Technology,
  Research Center J\"ulich GmbH (FZJ) \\
  D-52425 J\"ulich, Germany
}

\author{
  A.~Wallraff
  and
  A.~V.~Ustinov
}

\address{
  Physikalisches Institut III,
  Uni\-ver\-si\-t\"at Er\-lan\-gen-N\"urn\-berg,
  D-91054, Erlangen, Germany
}

\date{\today}

\wideabs{ 

\maketitle

\begin{abstract}

 We present a systematic study of the Cherenkov radiation of Josephson plasma waves by fast moving fluxon in a stack of coupled long Josephson junctions for different fluxon modes. It is found that at some values of parameters current-voltage characteristic may exhibit a region of the back-bending on the fluxon step. In the opposite limit the emission of the Cherenkov radiation takes place. In the annular junctions of moderate length the interaction of the emitted waves with fluxon results in the novel resonances which emerge on the top of the fluxon step. We present more exact formulas which describe the position of such resonances taking into account difference between junction and non-linear corrections. The possibility of direct detection of the Cherenkov radiation in junctions of linear geometry is discussed.

\end{abstract}

} 

\section{Introduction}

Cherenkov radiation exists if a particle moves with a velocity equal to the phase velocity of the emitted waves. Electromagnetic waves in a long Josephson junction (LJJ) are described by the sine-Gordon equation which has, in particular, soliton and plasma wave solutions. Physically, the solitons are Josephson fluxons (often called Josephson vortices), and plasma waves in most cases are just small amplitude linear electromagnetic waves. A soliton behaves here as a quasiparticle with its own characteristic mass and velocity. Therefore, one may consider the problem of Cherenkov radiation of plasma waves induced by fast moving solitons.

The dynamics of Josephson phase $\phi$ in a LJJ is rather accurately described by the perturbed sine-Gordon equation \cite{McLoughlinScott}:
\begin{equation}
      \phi_{\tilde x\tilde x} -  \phi_{\tilde t\tilde t} - \sin{\phi}
      =   \alpha \phi_{\tilde t} - \gamma
  \quad , \label{Eq:sG}
\end{equation}
where \rhs of (\ref{Eq:sG}) is usually referred as perturbation. The coordinate $\tilde{x}$ is normalized to the Josephson penetration depth $\lambda_J$, the time $\tilde{t}$ is normalized to the inverse plasma frequency $1/\omega_p$, $\gamma=j/j_c$ is the bias current density $j$ normalized to the critical current density $j_c$, $\alpha=1/\sqrt{\beta_c}$ is the damping coefficient, $\beta_c$ is the well known McCumber-Stewart parameter. Generally, it is assumed that $\alpha$ and $\gamma$ are small.

If we neglect the dissipation and the bias current [\rhs of (\ref{Eq:sG})] and consider a small amplitude wave
\begin{equation}
  \phi=A \sin(k\tilde{x}-\omega\tilde{t}) \quad , \quad A \ll 1
  \quad , \label{Eq:wave}
\end{equation}
by substituting (\ref{Eq:wave}) into (\ref{Eq:sG}) we obtain the well known dispersion relation for small amplitude plasma waves in LJJ:
\begin{equation}
  \omega = \sqrt{k^2 + 1}
  \quad . \label{Eq:LJJ:omega(k)}
\end{equation}
The phase velocity of these waves is
\begin{equation}
  u_{\rm ph} = \frac{\omega(k)}{k}
  = \sqrt{1+\frac{1}{k^2}}
  \quad , \label{Eq:u_ph}
\end{equation}
and can take any value between 1 and $\infty$. Here the velocity $u$ is normalized to the, so-called, Swihart velocity ${\bar c}_{0}=\omega_p\lambda_J$. From (\ref{Eq:u_ph}) it follows that the Swihart velocity is the minimum phase velocity of electromagnetic waves in the system.

The fluxon moving with the velocity $u$ in a LJJ can be described by the following solution of (\ref{Eq:sG}) without $\alpha$ and $\gamma$ terms:
\begin{equation}
  \phi = 4 \arctan \exp \left(
    \frac{\tilde{x}-u\tilde{t}}{\sqrt{1-u^2}}
  \right)
  \quad . \label{Eq:fluxon}
\end{equation}
Taking into account the \rhs of (\ref{Eq:sG}) as a perturbation \cite{McLoughlinScott} one may derive the force-balance dependence of the fluxon velocity $u$ on the bias current $\gamma$ and the dissipation coefficient $\alpha$:
\begin{equation}
  u = \frac{1}{\sqrt{1+\left( \frac{4\alpha}{\pi\gamma} \right)^2}}
  \quad . \label{Eq:u(gamma)}
\end{equation}
The allowed range of fluxon velocities lies between $0$ and $1$. Thus, the maximum fluxon velocity $u_{\max}$ coincides with the minimum possible phase velocity ${\bar c}_{0}$ of linear electromagnetic waves in the system and Cherenkov radiation can not take place. This coincidence results from the structure of sine-Gordon equation and, in principle, may not exist in more complex systems such as coupled LJJ's, arrays of JJ's or systems with special resonant structures \cite{KurinYulin}. Coupled systems received a lot of attention during the last few years due to the progress achieved in fabrication of Nb-Al-AlO$_x$-Nb low-$T_c$ tunnel junctions and studies of layered high-$T_c$ superconductors that exhibit intrinsic Josephson effect \cite{Intrinsic}.

The behavior of Josephson phases in a system of coupled LJJ's is described by a system of coupled sine-Gordon equations \cite{SBP}.
Since the dispersion relation for linear waves as well as the maximum velocity of a fluxon are influenced by mutual coupling between the junctions, one may find the conditions at which Cherenkov radiation appears in such a system. Recently we reported clear evidence \cite{Cherry1} of radiation by fluxons moving in a system of 2 coupled annular LJJ's. In this paper we present systematic study of the dependence of this effect on junction parameters, location of the trapped fluxon and report a more detailed model then the one presented in Ref.~\onlinecite{Cherry1}.

The paper is organized as follows. In section II we present a model which extends the model of Ref.~\onlinecite{Cherry1} in several aspects. The numerical technique and the results of simulations are presented in section III. The theory from section II is further extended to describe the position of resonanaces obtained in simulation more accurately. The question whether Cherenkov radiation can be observed directly in LJJ of linear geometry is discussed. Section IV shows experimental results. The phenomenon of Cherenkov radiation in $N$-fold ($N>2$) stack is discussed in section V. Section VI concludes the work.

\section{Model for two coupled junctions}

From both practical and theoretical point of view, it is interesting first to consider the most simple system such as two coupled (generally asymmetric) LJJ's, which we denote as LJJ$^A$ and LJJ$^B$ from now on. They can be described by two coupled perturbed sine-Gordon equations \cite{suk:94,LT21,Radio}:
\begin{equation}
  \left\{
    \begin{array}{rrr}
      \frac{1}{1-S^2} \phi^A_{\tilde x\tilde x}
      - & \phi^A_{\tilde t\tilde t}
      - & \sin{\phi^A}
      -   \frac{S\sqrt{D'}}{1-S^2} \phi^B_{\tilde x \tilde x}
      =   \alpha^{A} \phi^A_{\tilde t}
      -   \gamma^A
      \\
      \frac{D'}{1-S^2} \phi^B_{\tilde x\tilde x}
      - & \frac{1}{C}\phi^B_{\tilde t\tilde t}
      - & \frac{1}{J} \sin{\phi^B}
      -   \frac{S\sqrt{D'}}{1-S^2} \phi^A_{\tilde x\tilde x}
      =   \alpha^{B} \phi^B_{\tilde t}
      -   \gamma^B
    \end{array}
  \right.
  \quad , \label{EqsBasicNorm}
\end{equation}
where $S$ ($-1<S<0$) is a dimensionless coupling constant, $D'=d'^A/d'^B$ is the ratio of the effective magnetic thicknesses of LJJ$^A$ and LJJ$^B$, $C=C^A/C^B$ is the ratio of specific capacitances, $J=j_c^A/j_c^B$ is the ratio of critical currents, $\alpha^{A,B}$ and $\gamma^{A,B}=j^{A,B}/j_c^{A,B}$ are the damping coefficients and normalized bias currents, respectively.

\subsection{Plasma waves}

For the analysis of linear modes we use Eq.~(\ref{EqsBasicNorm}) in the absence of perturbations. Considering the small amplitude waves
\begin{equation}
  \phi^{A,B}=A^{A,B}\sin\left( k\tilde{x} - \omega \tilde{t} \right)
  \quad , \quad A \ll 1
  \quad , \label{Eq:SinWave}
\end{equation}
in such a system and linearizing $\sin\phi^{A,B}\approx\phi^{A,B}$ we obtain a dispersion relation which consists of two branches corresponding to two different modes of small-amplitude electromagnetic wave propagation in the system: in-phase mode and out-of-phase mode. In general, this relation is rather bulky\cite{Note:DispRelation} but for the case $D'=C=1$, which we will be mostly interested in, it reduces to the simpler form:
\begin{equation}
  \omega(k) = \sqrt{
    \frac{1+J}{2J} + \frac{k^2}{1-S^2} \pm
    \sqrt{
      \frac{\left( 1-J \right)^2}{4J^2} +
      \frac{S^2k^4}{\left( 1-S^2 \right)^2}
    }
  }
  \quad . \label{Eq:omega_pm(k)}
\end{equation}
From (\ref{Eq:omega_pm(k)}) the phase velocity $u_{\rm ph} = \omega(k)/k$ can be calculated as
\begin{equation}
  u_{\rm ph} = \sqrt {
    \frac{1+J}{2Jk^2} + \frac{1}{1-S^2} \pm
    \sqrt{
      \frac{\left( 1-J \right)^2}{4J^2k^4} +
      \frac{S^2}{\left( 1-S^2 \right)^2}
    }
  }
  \quad . \label{Eq:u_ph(k)}
\end{equation}
The phase velocities of the in-phase mode [``$+$'' sign in (\ref{Eq:omega_pm(k)}) and (\ref{Eq:u_ph(k)})] are in the range from $\bar{c}_{+}$ to $\infty$ and for the
out-of-phase mode [``$-$'' sign in (\ref{Eq:omega_pm(k)}) and (\ref{Eq:u_ph(k)})] from $\bar{c}_{-}$ to $\infty$. In the general case, the velocities $\bar{c}_{\pm}$ are defined as \cite{LT21}
\begin{equation}
  \bar{c}_{\pm}
  = \sqrt{\frac{1+D'C \pm \sqrt{(D'C-1)^2 + 4 D'C S^2}}{2 (1-S^2)}}
  \quad , \label{Eq:c_pm}
\end{equation}
and have the same meaning as the Swihart velocity ${\bar c}_{0}$ for a single LJJ but belong to different modes of electromagnetic wave propagation. In the case $D'=C=1$ the expression (\ref{Eq:c_pm}) for Swihart velocities reduces to:
\begin{equation}
  \bar{c}_{\pm} = \frac{1}{\sqrt{1 \pm S}}
  \quad . \label{Eq:Sym:c_pm}
\end{equation}
Note, that in the this case $\bar{c}_{\pm}$ do not depend on $J$.

In a single LJJ, the maximum fluxon velocity $u_{\max}$ coincides with the Swihart velocity of plasma waves ${\bar c}_{0}$. But in general, the Swihart velocities [\eg\ (\ref{Eq:c_pm})] have {\em nothing} in common with the possible range of fluxon velocities in the coupled system since no solution similar to (\ref{Eq:fluxon}) is known in general case.

\subsection{Fluxons}

The variety of fluxon configurations is very rich even for 2 coupled LJJ's. When discussing different fluxon configurations we will use the notation $[N|M]$, which means $N$ fluxons located (trapped) in
LJJ$^A$ and $M$ fluxons in LJJ$^B$ ($N,M<0$ describe
anti-fluxons).

{\em The $[1|0]$ state.}\/ Let's consider now two coupled LJJ with $J\ne1$ and some travelling wave solution $\phi^{A,B}(\tilde{x}-u\tilde{t})$. Introducing a new variable $\xi=\tilde{x}-u\tilde{t}$ and omitting $\alpha$ and $\gamma$ terms, the unperturbed Eqs.~(\ref{EqsBasicNorm}) can be rewritten as
\begin{equation}
  \left\{
    \begin{array}{rr}
      \phi^A_{\xi\xi} \left( \displaystyle{\frac{1}{1-S^2} - u^2} \right)
      - \displaystyle\frac{S\sqrt{D'}}{1-S^2} \phi^B_{\xi\xi}
      = & \sin{\phi^A}
      \\
      \phi^B_{\xi\xi}
      \left( \displaystyle{\frac{D'}{1-S^2} - \frac{u^2}{C}} \right)
      - \displaystyle\frac{S\sqrt{D'}}{1-S^2} \phi^A_{\xi\xi}
      = & \displaystyle\frac{1}{J}\sin{\phi^B}
    \end{array}
  \right.
  \quad . \label{Eq:x-ut}
\end{equation}
Let's now suppose that the solution $\phi^{A,B}$ moves with velocity $u={\bar c}_{-}$ (\ref{Eq:c_pm}). Substituting $u={\bar c}_{-}$ from (\ref{Eq:c_pm}) into (\ref{Eq:x-ut}) we obtain the following relation between $\phi^A$ and $\phi^B$:
\begin{equation}
  \sin\left( \phi^A \right) = \kappa \sin \left( \phi^B \right)
  \quad , \label{Eq:[1|0]@c_}
\end{equation}
where
\begin{equation}
  \kappa =
  \frac{D'C-1+\sqrt{(D'C-1)^2 + 4 D' C S^2}}
       {2SJ\sqrt{D'}}
  \quad . \label{Eq:kappa}
\end{equation}
In the simplest case $D'=C=1$ and $J\ne1$ we have:
\begin{equation}
  {\bar c}_{-} = \frac{1}{\sqrt{1-S}} \quad , \quad
  \kappa = \frac{1}{J}
  \quad . \label{Eq:kappa@D'=1}
\end{equation}

After establishing Eq.~(\ref{Eq:[1|0]@c_}), let's consider $[1|0]$ state with the only fluxon in LJJ$^A$, \ie, $\phi^A$ grows from $0$ to $2\pi$ as $x$ changes from $-\infty$ to $+\infty$ and $\phi^B$ changes with $x$ but $\phi^B(\pm\infty)=0$. If $\kappa=1$ then from Eq.~(\ref{Eq:[1|0]@c_}) it follows that $\phi^A=\phi^B + 2 \pi k$. Thus, if $\phi^A$ grows from 0 to $2\pi$, $\phi^B$ also grows in the same way which corresponds to the state $[1|1]$ and contradicts to the assumption that we consider the state $[1|0]$. Therefore in the state $[1|0]$ with $\kappa=1$ the fluxon can not move with the velocity ${\bar c}_{-}$ --- it is the maximum asymptotic (not reachable) velocity of fluxon.

If $\kappa>1$ and $\phi^A$ grows from $0$ to $2\pi$ as $x$ changes from $-\infty$ to $+\infty$, $\phi^B$ will not make $2\pi$-leap as $x$ changes from $-\infty$ to $+\infty$. Imagine that $\phi^A$ grows from $0$ and approaches $\pi/2$ at some point $x=x_0$, \ie, $\sin(\phi^A)$ approaches $1$. At this moment $\sin(\phi^B)$ will approach $1/\kappa$ ($\kappa>1$) and $\phi^B$ will be equal to some value between $0$ and $\pi/2$ (\eg\  $\pi/3$). At the next point $x=x_1>x_0$ where $\phi^A$ comes over $\pi/2$, $\sin(\phi^A)$ starts to decrease, so $\sin(\phi^B)$ becomes smaller than $1/\kappa$ and $\phi^B < \pi/3$, \ie, $\phi^B$ will not make a $2\pi$-leap. This is an important point: the fact that $\phi^B$ does not twist means that there is no fluxon in LJJ$^B$, \ie, $[1|0]$ state is possible at the velocity $u={\bar c}_{-}$.

From mathematical point of view (for a ``good'' differential equation), if a solution exists at point $u={\bar c}_{-}$ it will also exist in the vicinity of this point, \ie, also above ${\bar c}_{-}$. In fact, simulations show that for $D'=C=1$, $J=0.5$, $S=-0.5$ and $\alpha=0.1$ the state $[1|0]$ can survive up to about ${\bar c}_{0}$ at some parameters of the system. This region of velocities ${\bar c}_{-}<u<u_{\max}^{[1|0]}$ is a domain were Cherenkov radiation in $[1|0]$ state is excited by a moving fluxon. Of course $u_{\max}^{[1|0]}$ is a function of such parameters as $J$, $S$, $\alpha$.

For $\kappa<1$ a fluxon in the $[1|0]$ state can not reach the velocity ${\bar c}_{-}$. Moreover, the maximum velocity of the fluxon in this case is smaller than ${\bar c}_{-}$.

Here we have to stress that rather special conditions are required to get Cherenkov emission in this system. The most important condition is the {\em asymmetry} of the two-fold stack. For the $[1|0]$ state not only the condition $J<1$ leads to $u_{\max}>{\bar c}_{-}$ but also, \eg, $\gamma^B<\gamma^A$ results in a similar phenomenon as we checked numerically. If one includes $\gamma^A\ne\gamma^B$ into the coupled equations not as a perturbation but as initial offset of $\phi^B$, it is possible to obtain an equation very similar to the above equations for $J\ne1$.

{\em The $[1|\pm1]$ state.}\/ If LJJ's are identical ($C=D'=J={\alpha^A}/{\alpha^B}=1$), the coupled sine-Gordon system consists of two identical equations and, due to the symmetry, in the state $[1|\pm1]$ the relation $\phi^A=\pm\phi^B$ can be satisfied. Therefore, two Eqs.~(\ref{EqsBasicNorm}) reduce to one equation
\begin{equation}
  \frac{1}{1 \pm S} \phi_{\tilde{x}\tilde{x}}-
  \phi_{\tilde{t}\tilde{t}}-
  \sin\phi = \gamma - \alpha \phi_{\tilde{t}}
  \quad , \label{Eq:Degen_sG}
\end{equation}
which is the usual sine-Gordon equation with the Josephson penetration depth $\lambda_J^{\pm}=\sqrt{1 \pm S}$ (in normalized units) and the modified maximum velocity of fluxon $u_{\max}^{\pm}=1/\sqrt{1 \pm S}={\bar c}_{\pm}$. Since the fluxons in these modes can move with any velocity from 0 to ${\bar c}_{\pm}$, they can not excite in-phase plasma waves by means of the Cherenkov mechanism. The state $[1|-1]$ with the maximum velocity ${\bar c}_{-}$ can not excite any plasma wave mode at all. Only the fluxons in the state $[1|1]$ moving with velocity ${\bar c}_{-}<u<{\bar c}_{+}$ may, in principle, generate the out-of-phase mode of plasma waves. This may happen for two coupled LJJ's with different parameters. In the symmetric case $\phi^A=\phi^B$ Cherenkov radiation does not occur since the solution of Eq.~(\ref{Eq:Degen_sG}) is purely solitonic and symmetric while Cherenkov radiation requires the presence of the out-of-phase component.

If the two LJJ's are not identical, the pure degeneration of equations does not take place and Cherenkov radiation can be generated by two fluxons moving in $[1|1]$ state. In general, the derivation follows the one from Ref.~\onlinecite{GrE:Stability} but, in Eqs. (7)--(9) of Ref.~\onlinecite{GrE:Stability}, the factor 2 in front of the sine should be changed to $(J+1)/2J$ and the running coordinate to
\[
\xi = \frac{x-ut}{\sqrt{1-u^2+\Delta}} \sqrt\frac{1+J}{2J}.
\]

\section{Numerical simulation}

The simulations were performed using the system of Eqs.~(\ref{EqsBasicNorm}) with the simulation technique described in detail in Ref.~\onlinecite{Radio}. Investigating Cherenkov phenomena, we took additional care about discretization. For simulation we use a finite difference method in which continuous LJJ is represented as a discrete chain of point-like LJJ's with very small spacing $a \ll 1$. Since the dispersion relation for plasma waves in a discrete medium is qualitatively different from the continuous one for large $k$, we have to find a parameter range in which the discreteness effects can be ignored. The dispersion relation for the out-of-phase mode of linear plasma waves in the coupled discrete array of JJ is ($D'=C=J=1$) \cite{StkArrDispRel}:
\begin{equation}
  \omega(k)  = \sqrt{1+\frac{4}{a^2}\sin^2\left( \frac{ak}{2\sqrt{1-S}} \right)}
  \quad , \label{Eq:DispRel4array}
\end{equation}
\ie, it is periodic in $k$. Therefore, the fluxon dispersion line $\omega=uk$ intersects the dispersion relation (\ref{Eq:DispRel4array}) for any $u>0$, \ie, even for $u \le {\bar c}_{-}$ and, therefore, Cherenkov radiation can occur in simulation but not in a real continuous system. For $k \ll \pi/a$ the dispersion relation (\ref{Eq:DispRel4array}) is identical to that of the continuous system. Since in continuous system the radiation does not occur for fluxon velocities up to ${\bar c}_{-}$, let's calculate the wave vector $k_c$ of plasma wave which will be emitted in the discrete system at the fluxon velocity $u={\bar c}_{-}$. This value is determined by the intersection of the fluxon dispersion line $\omega={\bar c}_{-}k$ with the dispersion curve of plasma waves (\ref{Eq:DispRel4array}):
\begin{equation}
  {\bar c}_{-}k_c =
  \sqrt{1+\frac{4}{a^2}\sin^2\left( \frac{ak_c}{2\sqrt{1-S}} \right)}
  \quad . \label{Eq:find_k_c}
\end{equation}
Taking into account (\ref{Eq:Sym:c_pm}) we solve Eq.~(\ref{Eq:find_k_c}) for $k_c$. Since for $a \ll 1$ Eq.~(\ref{Eq:DispRel4array}) is very close to the dispersion relation of the continuous system in the range $k \ll\pi/a$, it is sufficient to take into account small nonlinearity of sine in (\ref{Eq:find_k_c}), \ie\  the cubic term. Solving (\ref{Eq:find_k_c}) for $k_c$ we obtain the following expression:
\begin{equation}
  k_c^2 = 2\sqrt{3}\frac{1-S}{a}
  \quad . \label{Eq:k_c}
\end{equation}
Thus, for a fluxon moving with velocity $u \le {\bar c}_{-}$, any radiation with wavelength $\lambda_c=2\pi/k_c$ and smaller is an artifact of the simulation procedure. All simulation results below were obtained with $a=0.025$ and $S=-0.5$ which gives $\lambda_c = 0.44$ (in units of $\lambda_J$).

\subsection{Annular junctions}

The fluxon dynamics in annular junctions was simulated using Eqs.~(\ref{EqsBasicNorm}) with periodic boundary conditions:
\begin{eqnarray}
  \phi  ^{A,B}(0,\tilde{t}) & = &\phi  ^{A,B}(\ell,\tilde{t}) + 2 \pi N^{A,B}\\
  \phi_{\tilde{x}}^{A,B}(0,\tilde{t}) &
  = &\phi_{\tilde{x}}^{A,B}(\ell,\tilde{t})
  \quad , \label{Eq:AnnBC}
\end{eqnarray}
where $N^{A,B}$ is the number of trapped fluxons in LJJ$^{A,B}$, respectively, and $\ell=L/\lambda_J^A$ is the normalized circumference of the stack.

{\em The state $[1|0]$.}\/ The family of $I$--$V$ characteristics (IVC's) of the system for $D'=C=1$ and different values of $J=0.5$, $1.0$, $2.0$ in $[1|0]$ state is shown in Fig.~\ref{Fig:FamilyIVC[1|0]}. The circumference of the stack $\ell=20$ and the damping coefficient $\alpha=0.2$ were chosen to emulate an infinite junction ($\alpha\ell\gg1$). As we predicted above, for $\kappa>1$ ($J<1$) the fluxon velocity exceeds ${\bar c}_{-}$ and Cherenkov radiation takes place. The phase gradient profiles for $u>{\bar c}_{-}$ are similar to those discussed below. For $\kappa<1$ ($J>1$) we find that $u_{\max}<{\bar c}_{-}$. Note that in this case there is a region of back-bending of IVC, after which LJJ$^B$ switches to the resistive state.

The presence of the back-bending on the IVC is a rather interesting phenomenon. Though, it is quite usual for non-linear resonances, in our case there is no real resonance. It seems not trivial also from physical point of view: we pump energy into the system by means of bias currents but the wave (fluxon in LJJ$^A$ and its image in LJJ$^B$) as a whole moves slower. By the word ``image''\cite{image} we call the perturbation of the Josephson phase which takes place in LJJ$^B$ due to the presence of the fluxon in LJJ$^A$. The difference between image and anti-fluxon is that anti-fluxon carries the flux $-\Phi_0$ and image carries no flux.

First, we suspected that the state of the system at the back-bending part of IVC is metastable and the system switches to some other state after a long period of time. However, even after waiting more than $10^4$ time units the system is still at the back bent region of the IVC and dc voltage on the LJJ (velocity of the fluxon) has the same value within the specified accuracy (better than $0.1\,{\rm \%}$). Since in the back-bending region the state is stable, we suspect that there should be another reason, \eg\ a structural transformation of the solution, which may cause the negative differential resistance.

We propose the following qualitative explanation based on numerical observations. At the point where $R_d=du/d\gamma$ approaches zero, a new strongly nonlinear solution --- a breather (fluxon-antifluxon pair) --- forms in LJJ$^B$. At the beginning, the distance between the fluxon and antifluxon in the breather is small and the amplitude of the breather is close to zero. The breather does not oscillate since it is stretched by the bias current $\gamma^B$ which pulls the fluxon and antifluxon in opposite directions. The higher the bias current, the stronger it stretches the breather, the larger is the distance between the fluxon and antifluxon. As the distance increases, the profile of the breather becomes similar to the profile of a fluxon and an antifluxon situated at some distance from each other. Therefore, as the distance between the fluxon and antifluxon grows from 0 to about $\lambda_J$, the friction force grows from 0 to about 2$F_{\alpha}$, where $F_{\alpha}$ is the friction force acting on the fluxon. Since the friction force grows very fast with $\gamma^B$, much faster than the driving force, this leads to a back-bending region on the IVC. At some point, the current $\gamma^B$ becomes large enough to break the fluxon-antifluxon pair situated in LJJ$^B$ and, after rather complicated transient process, LJJ$^B$ switches to the non-zero voltage state (R-state).

To learn about this scenario, we simulated the transient process of switching of LJJ$^B$ to the R-state for $J=2.0$. Starting from the point $\gamma^B=0.35$ at the very top of the back-bending part of the IVC, the current was increased by a small amount $\Delta\gamma=0.05$ which was enough for LJJ$^B$ to switch to the R-state. We observed a dissociation of the fluxon-antifluxon pair in LJJ$^B$. Three snapshots of the transient process of fluxon-antifluxon dissociation are shown in Fig.~\ref{Fig:FAF_Separation}.

Fig.~\ref{Fig:FamilyIVC[1|0]} represents the IVC of the long, resonance-free system. The finite length of the stack ($\alpha\ell \lesssim 1$) results in additional effects \cite{Cherry1}. As an example, we simulated the IVC of the stack with $D'=C=1$, $J=0.5$, $S=-0.5$, $\alpha=0.1$ and $\ell=7$ in the $[1|0]$ state. According to the results of the previous section, for this set of parameters $u_{\max}>{\bar c}_{-}$, and Cherenkov radiation has to appear at ${\bar c}_{-}<u<u_{\max}$. The IVC is shown in Fig.~\ref{Fig:SimIVC_Ann}. The choice of parameters was close to that of our experimentally studied sample (see below) except for the damping coefficient $\alpha=0.1$ which was taken larger than in the experiment in order to decrease the calculation time. For comparison, the IVC's for $S=0$, and $S=-0.5$, $J=1$ are shown in Fig.~\ref{Fig:SimIVC_Ann} as well.

First, from simulations we find that for the bias current $\gamma>\gamma^{*}$ (see Fig.~\ref{Fig:SimIVC_Ann}) the soliton velocity $u$ becomes larger than $\bar{c}_{-}$, in accordance with consideration given above. Second, there are two steps to the right from $\bar{c}_{-}$. To understand the nature of these steps we calculated the phase gradient profiles $\phi^{A,B}_{\tilde{x}}(\tilde{x})$ for various points of the IVC. If one increases $\gamma$ from $0$ to $1$, the fluxon dynamics develops in the following way. In the region $0<u<\bar{c}_{-}$ the fluxon motion is qualitatively well described by the perturbation approach ($|S| \ll 1$) \cite{image} and we find the fluxon in LJJ$^A$ and its image in LJJ$^B$. Their profiles are shown in Fig.~\ref{Fig:Profiles}(a) that corresponds to the point A on the IVC (see inset in Fig.~\ref{Fig:SimIVC_Ann}). As soon as $u$ exceeds $\bar{c}_{-}$, a Cherenkov radiation wake arises behind the moving soliton and its image as shown in Fig.~\ref{Fig:Profiles}(b) which corresponds to the point B on the IVC. With increasing soliton velocity, the wavelength of the radiation $\lambda=2\pi/k$ increases in agreement with (\ref{Eq:u_ph(k)}), and the amplitude and length of the wake quickly grow. The amplitude of the wake decays exponentially in time and space as can be seen from Fig.~\ref{Fig:Profiles}(c) corresponding to the point C on the IVC (see inset in Fig.~\ref{Fig:SimIVC_Ann}). At any point of the IVC the area under the soliton profile is $\int \phi^{A}_{\tilde{x}}\,d{\tilde{x}}=2\pi$ and for the image $\int \phi^{B}_{\tilde{x}}\,d{\tilde{x}}=0$.

Since the soliton moves in the annular LJJ of finite length $\ell$, at some velocity $u$ the Cherenkov radiation wake extends over the length $\sim \ell$ so that the soliton and its image ``see'' their own radiation wakes after turning around the junction. This is illustrated in Fig.~\ref{Fig:Profiles}(d) and corresponds to point D on the IVC (see the inset of Fig.~\ref{Fig:SimIVC_Ann}). The interaction of the soliton with its Cherenkov radiation wake results in the appearance of resonances on the IVC of the system at $u>\bar{c}_{-}$. We call these resonances ``Cherenkov steps'' (CS's). At the resonance the oscillations of $\phi^{A,B}$ in the LJJ's induced by the moving soliton take place in phase (show constructive interference) with Cherenkov-generated Josephson plasma waves $\phi^{A,B}$ [Fig.~\ref{Fig:Profiles}(d)]. Constructive interaction is possible for both fluxon and image at the same time since they, like out-of-phase plasma waves, have opposite polarity. If the fluxon moves in the minimum of the plasma wave in LJJ$^A$, the image moves in the maximum of plasma wave in LJJ$^B$ and such a state is the minimum energy state for both the fluxon and the image. Thus, the steps CS$_{2,3}$ on the IVC in Fig.~\ref{Fig:SimIVC_Ann} are related to the interference between the moving soliton and its Cherenkov radiation. Since the amplitude of the emitted plasma waves decays exponentially $\sim\exp(-\alpha \tilde{x})$, the condition of having resonances is $\alpha\ell\lesssim1$. In a long system or in the system with substantial dissipation this condition is not satisfied, the interaction of a fluxon with its wake does not occur and IVC looks smooth as shown in Fig.~\ref{Fig:FamilyIVC[1|0]} for $J=0.5$. Our numerical study showed that for $\alpha=0.1$ individual resonances appear up to $\ell\sim 30$. Looking at Fig.~\ref{Fig:FamilyIVC[1|0]} one can see that for $\alpha=0.2$ and $\ell=20$ resonances are not visible.

To calculate the positions of the resonances we suppose that resonance occurs if there is an integer number of plasma wavelengths over the length of the junction, \ie, the wave vector $k$ takes one of the following eigenvalues:
\begin{equation}
  k_m = \frac{2\pi}{\ell} m \mbox{ , where } m=1,2, \ldots , \infty.
  \label{Eq:k_m}
\end{equation}
Thus, Cherenkov radiation should lead to resonances at vortex
velocities equal to the phase velocity (\ref{Eq:u_ph(k)}) with $k$ given by (\ref{Eq:k_m}):
\begin{equation}
  u_m = \sqrt {
    \frac{1+J}{2Jk_m^2} +
    {\bar c}_{-}^2
    \frac{1+S \sqrt{\displaystyle
      \frac{\left( 1-J \right)^2\left( 1-S^2 \right)^2}
      {4J^2S^2k_m^4} + 1}}{1+S}
  }
  \quad . \label{Eq:u_m}
\end{equation}
The resonances calculated using (\ref{Eq:u_m}) are shown in Fig.~\ref{Fig:SimIVC_Ann} by thin vertical lines. The density of the resonances increases with $m$ up to infinity when $u\rightarrow\bar{c}_{-}+0$. Since the resonances have a finite width due to damping, it is possible to resolve only the CS's with not very large $m$ in LJJ's of moderate length $\ell$. The positions of resonances obtained from simulation (Fig.~\ref{Fig:SimIVC_Ann}) differ from the ones calculated using Eq.~(\ref{Eq:u_m}). We suggest several reasons for this discrepancy. First, it may result from the large amplitude of the waves which become strongly nonlinear at the resonance. Second, the gap in the dispersion relation decreases because of non-zero bias current $\gamma$. To take into account these two factors we write down a generalization of Eq.~(\ref{Eq:SinWave}) as
\begin{equation}
  \phi^{A,B}(\tilde{x},\tilde{t})
  =\phi_0^{A,B}
  + A^{A,B}\sin(k\tilde{x} - \omega \tilde{t})
  \quad , \label{Eq:SinWave+Support}
\end{equation}
and substitute it into Eq.~(\ref{EqsBasicNorm}) without $\alpha$-term. For further calculation we will exploit the following expansions \cite{Abramowitz&Stegun}:
\begin{equation}
  \sin \left( A \sin z \right) =
  2 \sum_{k=0}^{\infty} \BesselJ_{2k+1}(A) \sin (2k+1)z
  \quad , \label{Eq:sinAsin}
\end{equation}
and
\begin{equation}
  \cos \left( A \sin z \right) =
  \BesselJ_0(A) +
  2 \sum_{k=1}^{\infty} \BesselJ_{2k}(A) \sin 2kz
  \quad . \label{Eq:cosAsin}
\end{equation}
We are interested only in the dc and the first ac spectral component of these expansions. Expanding $\sin(\phi^{A,B})$ as a sine of a sum, we use the first harmonic of (\ref{Eq:sinAsin}) and dc component of (\ref{Eq:cosAsin}) to obtain the dispersion relation for the first harmonic of the large amplitude nonlinear plasma wave. This dispersion relation now depends on the amplitude of the plasma wave $A$ and the bias current $\gamma$. The resonance velocities are obtained as:
\begin{equation}
  u_m = \sqrt {
    \frac{1}{1-S^2} + \frac{P+Q}{2k_m^2} -
    \sqrt{\frac{(P-Q)^2}{2k_m^2} + \frac{S^2}{(1-S^2)^2}}
  }
  \quad , \label{Eq:u_m:NL}
\end{equation}
where
\begin{equation}
  P = \frac{2\BesselJ_1(A)}{A}
  \sqrt{1-\left[ \frac{\gamma  }{\BesselJ_0(A)} \right]^2}
  \quad , \label{Eq:P}
\end{equation}
\begin{equation}
  Q = \frac{2\BesselJ_1(A)}{JA}
  \sqrt{1-\left[ \frac{\gamma J}{\BesselJ_0(A)} \right]^2}
  \quad , \label{Eq:Q}
\end{equation}
The average amplitude of the plasma waves $A$ is assumed to be the same in both LJJ's. Expression (\ref{Eq:u_m:NL}) implies that resonant velocities $u_m$ decrease as $\gamma$ or $A$ increase. This means that CS's can be bent back since both $\gamma$ and $A$ increase as the bias point moves up along the CS. In real systems this behavior is compensated by damping and resonances may or may not be bent back.


We would like to mention that the simplified expression used in our previous work \cite{Cherry1} described positions of resonances rather well despite that nonlinearity of the waves, $\gamma\ne0$ and $J\ne1$ were not taken into account. This coincidence occurred because the effect of $J<1$ and the effect of the nonlinearity with $\gamma>0$ have opposite signs and, for our particular parameters, nearly cancel each other.

The appearance of the CS's is rather convincing indication of the Cherenkov radiation. The steps appear due to the constructive interaction between the fluxon and the image from one side and the out-of-phase plasma wave from the other side. The experimental observation of the CS's was accomplished for the first time by our group \cite{Cherry1} and is described in the next section.

{\em The state $[1|1]$.}\/ As it was mentioned in the previous section, if two LJJ's are identical the $[1|1]$ mode does not emit any radiation. We accomplished a series of simulations to investigate how the spread in parameters affects the motion of the fluxon. Examples of numerically calculated fluxon profiles are shown in Fig.~\ref{Fig:CherryIn[1|1]}. All parameters of LJJ's were identical except for the damping coefficients $R=1.5$ in Fig.~\ref{Fig:CherryIn[1|1]}(a) and critical currents $J=1.5$ Fig.~\ref{Fig:CherryIn[1|1]}(b). We found that $J\ne1$ as well as $R\ne1$ result in Cherenkov emission of out-of-phase plasma waves. Increasing the parameter spread one decreases the stability of the in-phase fluxon step and, at some spread, the $[1|1]$ state is destroyed and the in-phase fluxon step disappears. In the $[1|1]$ state, the interaction of the fluxon and the image with the out-of-phase plasma wave can not be constructive since two fluxons have the same polarity while the plasma waves have the opposite one. Thus, if the condition of minimum energy is held for LJJ$^A$, it is not held for LJJ$^B$. As a result, the IVC has usual smooth behavior without additional resonances. Therefore, experimental observation of Cherenkov radiation from fluxons moving in the $[1|1]$ state is a rather difficult task.

\subsection{Linear junctions}

The fluxon dynamics in linear junctions was simulated using Eqs.~(\ref{EqsBasicNorm}) with boundary conditions which correspond to zero applied external magnetic field:
\begin{eqnarray}
  \phi_{\tilde{x}}^{A,B}(0,\tilde{t})
  = \phi_{\tilde{x}}^{A,B}(\ell,\tilde{t}) = 0
  \quad . \label{Eq:LinBC}
\end{eqnarray}

In stacks of linear geometry (open ends) one may also observe Cherenkov radiation and resonances similar to that in annular stacks. In this case the fluxon moves at zero applied magnetic field reflecting as antifluxon from the edges of the junction. In the $[1|0]$ state, the image has also to change its polarity upon reflection. This motion is similar to the fluxon motion in single LJJ biased on zero-field step (ZFS). At a fluxon velocity larger than ${\bar c}_{-}$, the Cherenkov radiation wake appears. When the length of the wake becomes comparable with the length of the junction, the Cherenkov emission forms a standing plasma wave. For the case of the $[1|0]$ state the interaction between the standing plasma waves and the fluxon is constructive and leads to the series of resonances (CS's) shown in  Fig.~\ref{Fig:SimIVC_Lin}. The fluxon dynamics at the CS is rather similar to the fluxon dynamics in a single LJJ at the Fiske Step (FS), where moving fluxons interact with standing cavity mode. In the case of Fiske steps, the plasma wave is excited due to boundary conditions --- when the fluxon leaves the junction, a plasma wave is emitted. In the case of coupled LJJ's the energy is pumped into standing plasma wave by means of Cherenkov mechanism. The positions of the resonances can be calculated using the Eq.~(\ref{Eq:u_m:NL}) taking $2\ell$ as the length of the junction in the expression for $k_m$ (\ref{Eq:k_m}). We have to mention that reflection of the image results in a large energy dissipation that decreases the stability of the ZFS $[1|0]$ mode (height of ZFS) as the coupling $|S|$ or the number of coupled junctions $N$ increases.

For the junctions of linear geometry one may try to detect the Cherenkov radiation directly by measuring the rf voltage at the junction boundary. Since the plasma wave oscillations are out-of-phase in the two LJJ's, it will be almost impossible to detect anything meaningful experimentally if the rf voltage is measured across the whole structure. Therefore, the sample should be designed in such a way that it allows to take the rf signal from one of the junctions.

At the working point chosen on a CS, the frequency of the plasma wave is a multiple of the Josephson frequency $\omega_{pl}=m\omega_J$, where $m$ is a resonance number. The spectrum of the voltage oscillation (fluxon part plus plasma wave part) in each junction will have a higher amplitude of the $m$-th harmonic. Nevertheless, such a behavior of $m$-th harmonic can not be considered as doubtless indication of Cherenkov emission. Simulation shows that in the $[1|0]$ state even the system with $J=1$ which should not exhibit Cherenkov effect has enhanced amplitudes of high harmonics. The power is pumped into high harmonics because of reflection of images at the edge of the junction when the plasma waves are emitted in a fashion similar to Fiske steps. Since this plasma wave interacts with the fluxon, one may observe resonances on the IVC of the system even at $u<{\bar c}_{-}$ as shown by point A in Fig.~\ref{Fig:SimIVC_Lin}. We will not consider this resonance here since it is a subject of separate investigation.

Thus, in the two-fold stack of linear geometry neither the presence of resonances on the IVC nor enhanced amplitude of high harmonics in the spectra can be a reliable indication of the Cherenkov emission. An example of the spectra that correspond to the points A, B, C and D on the IVC (Fig.~\ref{Fig:SimIVC_Lin}) is shown in Fig.~\ref{Fig:SpectraLin}.

\section{Experiment with two-layer Josephson ring}

Experiments have been performed with (Nb-Al-AlO$_{x}$)$_2$-Nb annular stacks. Due to the magnetic flux quantization in a superconducting ring, the number of vortices initially trapped in an annular junction is conserved. The soliton dynamics can be studied here under periodic boundary conditions, which exclude a complicated interaction of the soliton with the junction edges. The mean diameter of the stacked LJJ's was $D=132\,{\rm{\mu m}}$ and the width $W=10\,{\rm{\mu m}}$. The thickness of the middle Nb layer was about $60\,{\rm nm}$. The normalized circumference of the ring at $T=4.2\,{\rm K}$ was $\pi D/\lambda_J=L/\lambda_J\approx 7$, where $\lambda_J$ is the Josephson penetration depth. Fluxon trapping was realized by applying a small bias current to the junction while cooling the sample through the critical temperature ${T_c}^{\rm Nb}=9.2\,{\rm{K}}$. Measurements were performed at $4.2\,{\rm{K}}$.

In stacked annular LJJ's, trapping of a single fluxon is rather difficult due to the asymmetry of the required state $[1|0]$. Therefore, after every trapping attempt, the resulting state was carefully checked by several means. First, the IVC of the system was traced. A large critical current $I_c$ implies the $[0|0]$ configuration. If $I_c$ is close to zero, the voltage of a soliton step on IVC has to be proportional to the total number of vortices trapped in the system minus the number of pinned vortices. Second, the dependencies of the critical current $I_c$ and the maximum current at the single junction gap voltage $I_g$ on magnetic field $H$ applied in the plane of the sample were measured. The soliton configuration was deduced from the shape of $I_c(H)$ and $I_g(H)$ \cite{Vernik:Ic(H)-Ann}. As shown in Fig.~\ref{Fig:ExpIc(H)}, in the $[1|0]$ state $I_c(H)$ has a minimum at $H=0$ and $I_g(H)$ has a maximum. The LJJ in the stack under investigation had the ratio of critical current densities $J=j^A_c/j^B_c$ of about $0.5$, therefore it was easy to distinguish which junction contains the fluxon.

The fluxon step in the $[1|0]$ configuration is shown in Fig.~\ref{Fig:ExpIVC}. In addition to the main fluxon step, we find two additional steps denoted CS$_2$ and CS$_3$ at voltages $V_{CS_2}\approx 33.3\,{\rm{\mu V}}$, and $V_{CS_3}\approx 30.3\,{\rm{\mu V}}$. Similar IVC's were observed in several other stacked annular LJJ's.

To find the experimental values for the velocities $\bar{c}_{+}$ and $\bar{c}_{-}$ in the stack, we measured the FS voltage spacing for both in-phase and out-of-phase modes \cite{Ust-stack:93} by applying a magnetic field in the plane of the sample. Using fields up to $35\,{\rm Oe}$ we observed two families of FS's at voltages above the single-junction gap voltage. In this case, one junction is biased at the gap voltage state and the excess voltage is generated by cavity resonances in the other junction. We obtained $\Delta V_{-}\approx 30\,{\rm \mu V}$ and $\Delta V_{+}\approx 56\,{\rm \mu V}$ for the FS spacing in the out-of-phase and in-phase mode, respectively. From this data the coupling parameter is estimated as
\[
  S = - \frac{\Delta V_{+}^2-\Delta V_{-}^2}{\Delta V_{+}^2+\Delta
  V_{-}^2} \approx - 0.55
\]
and the limiting velocities $\bar{c}_{+}\approx 0.038 c$,
$\bar{c}_{-}\approx 0.020 c$ and $\bar{c}_{0}\approx 0.025 c$ are
calculated. The voltage corresponding to the velocity ${\bar c}_{0}$ is equal to $V_0\approx 37.4\,{\rm{\mu V}}$. The steps CS$_{2,3}$ appear at vortex velocities above $\bar{c}_{-}$, and, therefore, correspond to the Cherenkov steps that are also found in numerical simulation (see Fig.~\ref{Fig:SimIVC_Ann}). The parameters for simulation were chosen by purpose very close to the parameters of our sample.

If the fluxon is trapped in another LJJ of the stack, the situation $J>1$ takes place and we could see the a small region on the top of the IVC which was bent back as shown in Fig.~\ref{Fig:ExpIVC-BB}. This bending back looks very similar to the one found in numerical simulation and shown in Fig.~\ref{Fig:FamilyIVC[1|0]}. Thus, experimental data show good agreement with numerical simulations for the both cases ($J<1$ and $J>1$) discussed above.

\section{Stacks with number of layers $N>2$}

In the ideal case of high-$T_c$ layered superconductors the intrinsic LJJ's have very similar critical current densities. The system of equations which describes $N$ inductively coupled LJJ's was derived in Ref.~\onlinecite{SBP} and in the case of Resistively Shunted Junction (RSJ) model and absence of perturbations (without $\alpha$ and $\gamma$ terms, \ie\ quasiparticle and bias currents) can be written as
\begin{equation}
  {\vec \phi}_{xx}
  = {\bf S} \sin {\vec \phi} + {\bf S} {\vec \phi}_{tt}
  \quad , \label{Eq:NsG}
\end{equation}
where
\begin{equation}
  {\bf S} = \left(
    \begin{array}{ccccccc}
      1 & S      & 0      & 0      & 0      & 0      & 0 \\
      S & 1      & S      & 0      & 0      & 0      & 0 \\
      0 & \ddots & \ddots & \ddots & 0      & 0      & 0 \\
      0 & 0      & S      & 1      & S      & 0      & 0 \\
      0 & 0      & 0      & \ddots & \ddots & \ddots & 0 \\
      0 & 0      & 0      & 0      & S      & 1      & S \\
      0 & 0      & 0      & 0      & 0      & S      & 1
    \end{array}
  \right)
  \label{Eq:S-matrix}
\end{equation}
is the matrix which describes the interaction between LJJ's, $-0.5<S \leq 0$ is a dimensionless coupling constant which for $N$-fold case can be expressed as:
\begin{equation}
  S =   \frac{-\lambda}
  {\displaystyle d_I\sinh\frac{d}{\lambda}+2\lambda\cosh\frac{d}{\lambda}}
  \quad , \label{Eq:S4N-fold}
\end{equation}
where $d_I$ is the thickness of insulator (tunnel barrier) between adjacent superconducting layers, $d$ is the thickness of each superconducting layer and $\lambda$ is the London penetration depth. In the limit of high-$T_c$ superconductors $d_I, d \ll \lambda$ and
\begin{equation}
  S\approx-\frac{1}{2}\left( 1-\frac{d_I d}{\lambda^2} \right)
  \quad . 
\end{equation}

\subsection{Plasma Waves}

It was derived that in an $N$-fold stack of LJJ's there are $N$ modes of plasma waves and, accordingly, $N$ dispersion branches with $N$ characteristic velocities. In the case of equal LJJ's parameters, the expression for $n$-th characteristic velocity is fairly simple \cite{Kleiner2D,suk:94}:
\begin{equation}
  \bar{c}_n^{(N)}
  = \frac{1}{\sqrt{1+2S \cos \left( \displaystyle\frac{\pi n}{N+1} \right)}}
  \quad , \quad
  (n=1 \ldots N).
  \label{SymStk:c_n^(N)}
\end{equation}
Note, that for given $N$ $\bar{c}_N^{(N)} < \ldots < \bar{c}_1^{(N)}$. As soon as a fluxon is accelerated up to the velocity $u>\bar{c}_N^{(N)}$, \ie\ above the lowest Swihart velocity, Cherenkov emission of the slowest mode with characteristic velocity $\bar{c}_N^{(N)}$ may take place. If $u$ exceeds several lowest Swihart velocities, the Cherenkov radiation may consist of a mixture of plasma waves of different modes with different $k$ and $\omega$ as illustrated in Fig.~\ref{Fig:Cherry_N-fold}. An example of numerical simulation of a Cherenkov radiation wake for $N=7$ is given in Ref.~\onlinecite{HexFisherCherry}.

\subsection{Fluxons}

In the case of two coupled LJJ's, the asymmetry is essential for Cherenkov emission to appear. An important question which arises for the case of $N$-fold stack, and especially for layered high-$T_c$ materials, is whether Cherenkov radiation may appear in a system where all junctions have the same parameters. In the case $N\ge3$ there is additional decrease of the symmetry of the system in comparison with the case $N=2$. For $N=2$ and $J=1$ both LJJ's are at the same conditions and each of them interacts with its neighboring junction in the same way. But already for $N=3$, the middle LJJ$^B$ has two neighbors while LJJ's$^{A,C}$ have only one. Let us take 3 coupled sine-Gordon equations and assume the $[0|1|0]$ or $[1|0|1]$ state with the symmetric solution $\phi^A=\phi^C$. In this case one can reduce 3 equations to two equations, one for $\phi^B$ and another for $\phi^A=\phi^C$:
\begin{equation}
  \begin{array}{lllll}
    \frac{1}{1-2S^2} \phi_{xx}^{A,C}
    &-\phi_{tt}^{A,C} &-\sin\phi^{A,C}
    &- \frac{S}{1-2S^2}\phi_{xx}^{B}
    &= \gamma - \alpha \phi_t^{A,C}
    \quad , \\[7pt]
    \frac{1}{1-2S^2} \phi_{xx}^{B}
    &-\phi_{tt}^{B} &-\sin\phi^{B}
    &- \frac{2S}{1-2S^2}\phi_{xx}^{A,C}
    &= \gamma - \alpha \phi_t^{B}
  \end{array}
  \label{Eq:Sym3fold}
\end{equation}
In Eq.~(\ref{Eq:Sym3fold}) for $\phi^B$ there is a factor $2$ in front of the coupling term. Thus, the problem of $3$-fold symmetric stack is reduced to the problem of $2$-fold asymmetric stack. Thus, the asymmetry appears automatically when one goes from $N=2$ to $N=3$.

The analysis presented above for the case $J\ne1$ can not be directly applied for asymmetric coupling term as in Eq.~(\ref{Eq:Sym3fold}). We simulated Eq.~(\ref{Eq:Sym3fold}) for the case of $[0|1|0]$ and $[1|0|1]$ fluxon states and found that the state $[0|1|0]$ is similar to the case $J<1$, \ie, the IVC bends to the right in the region $u>{\bar c}_{3}^{(3)}$ which results in emission of Cherenkov radiation, while the state $[1|0|1]$ is similar to the case $J>1$ and results in the IVC with the back-bending region at $u<{\bar c}_{3}^{(3)}$. Qualitatively, the IVC's for this case look like IVC's shown in Fig.~\ref{Fig:FamilyIVC[1|0]} for $J=0.5$ and $J=2.0$. Using this result, we predict that a fluxon moving in a layered high-$T_c$ superconducting material causes Cherenkov radiation and its IVC will not have the conventional single junction asymptotic behavior.

Since $N$-fold stacks with $N>2$ have lower symmetry, $[1|\ldots|1]$ fluxon state will emit Cherenkov radiation even in the case of equal LJJ's parameters. If the emission is very strong (for $N \gg 2$), the state $[1|\ldots|1]$ can not survive. We simulated three-fold stack in symmetric fluxon configuration $[1|1|1]$ and found that this state is still stable and emits Cherenkov radiation for $u>{\bar c}_{3}^{(N)}$ as shown in Fig.~\ref{Fig:111}.

In an $N$-fold stack of {\em linear geometry}, it turned out to be rather  difficult to stabilize the motion of fluxon in $[0|..|0|1|0|..|0]$ state in ZFS-like mode. Due to the presence of images in the neighboring (relative to the LJJ's with fluxon) LJJ's, the reflection conditions become not perfect (energy dissipates in collision of images and anti-images at the boundary) and the fluxon tends to leave the junction. The ideal reflection conditions are realized in uncoupled LJJ, \ie\ at $S=0$. When $|S|$ and/or $N$ (more images induced by one fluxon) increase, more energy dissipates during the reflection process. For a given $N$, at some critical value of $S$, the fluxon can not be reflected back and leaves the system. For layered high-$T_c$ superconductors ($N \gg 2$) $|S|$ approaches its maximum theoretical value of $0.5$. Thus, ZFS-like fluxon motion can most probably not be realized in the $[0|..|0|1|0|..|0]$ state in a high-$T_c$ layered stack with open boundaries. Therefore, Cherenkov radiation from a {\em single} fluxon moving in a stack of intrinsic LJJ's or in an $N$-fold stack ($N \gg 2$) can be studied only in an artificially made annular structure.

In experiment, one often deals with many rather than with only one fluxon moving in, \eg\ flux-flow mode. Will Cherenkov radiation appear in this case? Assuming $[0|\ldots|0|M|0|\ldots|0]$ state in the limit of dense uniform fluxon chain and neglecting the junction edges it is possible to show that Cherenkov radiation does not take place since the maximum velocity of chain in each mode is equal to ${\bar c}_{n}^{(N)}$. However, at high velocities when the fluxon chain becomes sparse, we may expect a crossover to the single fluxon behavior discussed above. The Cherenkov radiation will result in a modified structure of the top part of the flux-flow branch of the IVC.

\section{Conclusion}

We have shown that a single fluxon moving in a stack of long Josephson junctions can emit a Cherenkov radiation wake which consists of out-of-phase plasma waves. To generate Cherenkov radiation in the two-fold stack the proper asymmetry is required. In the $N$-fold ($N>2$) stack asymmetry is not necessary and a fluxon moving faster than the lowest Swihart velocity will emit Cherenkov radiation. This result is of great importance for the physics of high-$T_c$ materials which exhibit intrinsic Josephson effect.

For some fluxon configurations (\eg\ $[1|0]$), in a system of moderate length ($\alpha\ell\lesssim3$) the constructive interaction of the fluxon with the Cherenkov wave results in novel resonances emerging on the IVC. These resonances were observed in experiment with annular two-fold stack and their positions are calculated theoretically using more rigorous approach then in our previous work \cite{Cherry1}. It is shown that the direct observation of the Cherenkov radiation from the LJJ of linear geometry is problematic.

\acknowledgments

We would like to thank P.~Barbara, Ya.~Greenberg and V.~V.~Kurin for useful discussions, and N.~Thyssen for sample fabrication.



\begin{figure}
  \caption{
    The family of IVC's of two inductively coupled LJJ's for different values of $J$ in $[1|0]$ state. For $J<1$, $u_{\max}$ exceeds ${\bar c}_{-}$, for $J>1$, $u_{\max}<{\bar c}_{-}$.
  }
  \label{Fig:FamilyIVC[1|0]}
\end{figure}
\begin{figure}
  \caption{
    Snapshot of the transient process of switching of the LJJ$^B$ into R-state for $J=2.0$. The current $\gamma^B$ is large enough to break the breather.
  }
  \label{Fig:FAF_Separation}
\end{figure}
\begin{figure}
  \caption{
    Calculated dependence of velocity $u$ on bias current density
    $\gamma$ (normalized IVC) for one soliton in a 2-junction ring
    for $L/\lambda_J=7$, $\alpha=0.1$, $J=0.5$, $S=-0.5$. The inset shows
    in detail the top part of the IVC. The IVC's for different
    configurations are shown for comparison: dotted line corresponds to
    single annular junction ($S=0$), thin line to the stack in the $[1|-1]$
    state.
  }
  \label{Fig:SimIVC_Ann}
\end{figure}
\begin{figure}
  \caption{
    Profiles of $\phi^{A,B}_{\tilde{x}}(x)$ in
    both LJJ's. (a), (b), (c), (d) correspond, respectively, to IVC
    points A, B, C, D shown in the inset of Fig.~\ref{Fig:SimIVC_Ann}.
  }
  \label{Fig:Profiles}
\end{figure}

\begin{figure}[!htb]
  \caption{
    Cherenkov radiation of out-of-phase plasma waves generated by two fluxons moving in $[1|1]$ state in annular two-fold stack at $\gamma=0.6$, $S=-0.5$, $\ell=20$. (a) $R=1.5$, $u=1.27$; (b) $J=1.5$, $u=1.34$.
  }
  \label{Fig:CherryIn[1|1]}
\end{figure}

\begin{figure}
  \caption{
    The first ZFS with Cherenkov steps on the top of it on IVC of linear two-fold stack.
  }
  \label{Fig:SimIVC_Lin}
\end{figure}

\begin{figure}
  \caption{
    Spectra of rf voltage on LJJ$^A$,  on LJJ$^B$ and on both junction (black, white and gray) for different bias points are shown in plots (a)--(c), respectively, and correspond to the bias points A ($\gamma=0.16$), B ($\gamma=0.2$), C ($\gamma=0.35$) and D ($\gamma=0.53$) on the IVC shown in Fig.~\protect\ref{Fig:SimIVC_Lin}
  }
  \label{Fig:SpectraLin}
\end{figure}

\begin{figure}
  \caption{
    Dependence of the critical current $I_c$ and the maximum current at the gap voltage $I_g$ of the magnetic field $H$. The shape of these characteristics implies $[1|0]$ fluxon state.
  }
  \label{Fig:ExpIc(H)}
\end{figure}

\begin{figure}
  \caption{
    Experimental IVC of the double-layer annular junction in the $[1|0]$ soliton configuration for $J\approx0.5$. The Cherenkov resonances are marked CS$_{2,3}$. The inset shows the geometry of the sample.
  }
  \label{Fig:ExpIVC}
\end{figure}
\begin{figure}
  \caption{
    IVC of the double-layer annular junction in the $[1|0]$ soliton configuration for $J\approx2.0$. The back-bending region is visible.
  }
  \label{Fig:ExpIVC-BB}
\end{figure}
\begin{figure}
  \caption{
    Dispersion relation of $7$-fold stack with coupling constant $S=-0.5$ and dispersion line of a fluxon. At velocity $u>{\bar c}^{(N)}_{n-3}$ 3 modes of plasma waves are emitted.
  }
  \label{Fig:Cherry_N-fold}
\end{figure}
\begin{figure}
  \caption{
    Cherenkov radiation of out-of-phase plasma waves generated by fluxons moving in the $[1|1|1]$ state in an annular three-fold symmetric stack at $\gamma=0.6$, $S=-0.5$, $\ell=20$, $u=1.71$ (${\bar c}_{+}=1.84$).
  }
  \label{Fig:111}
\end{figure}


{\small

\begin{thebibliography}{99}

\bibitem[*]{email}
  e-mail: gold@hitech.cplire.ru


\bibitem{McLoughlinScott}
  D.~W.~McLaughlin, A.~C.~Scott,
  Phys. Rev.~A {\bf 18}, 1652 (1978).

\bibitem{KurinYulin}
  V.~V.~Kurin, and A.~V.~Yulin,
  Phys. Rev. B, {\bf 55}, 11659 (1997).

\bibitem{Intrinsic}
  R.~Kleiner, F.~Steinmeyer, G.~Kunkel, and P.~M\"{u}ller,
  Phys. Rev. Lett. {\bf 68}, 2394 (1992);\\
  %
  R.~Kleiner and P.~M\"{u}ller,
  Phys. Rev.~B {\bf 49}, 1327 (1994).

\bibitem{SBP}
  S.~Sakai, P.~Bodin, and N.~F.~Pedersen.
  J.~Appl. Phys. {\bf 73}, 2411 (1993).

\bibitem{Cherry1}
  E.~Goldobin, A.~Wallraff, N.~Thyssen, and A.~V.~Ustinov,
  Phys. Rev. B {\bf 57}, 130 (1998).

\bibitem{suk:94}
  S.~Sakai, A.~V.~Ustinov, H.~Kohlstedt,
  A.~Petraglia, and N.~F.~Pedersen,
  Phys. Rev.~B {\bf 50}, 12905 (1994).

\bibitem{LT21}
  E.~Goldobin, A.~Golubov, A.~V.~Ustinov,
  Czech. J. Phys. {\bf 46}, 663 (1996), LT-21 Suppl. S2

\bibitem{Radio}
  A.~Wallraff, E.~Goldobin, A.~V.~Ustinov,
  J.~Appl. Phys. {\bf 80}, 6523 (1996).

\bibitem{Note:DispRelation}
  It was obtained implicitly in Ref.~\onlinecite{Krasnov:10-Analytics}. To obtain explicit equation for $\omega(k)$ one has to substitute Eq.~(20) and Eq.~(21) of Ref.~\onlinecite{Krasnov:10-Analytics} into Eq.~(44) and use the fact that $u=\omega/k$. The explicit equation is given by Eq.~(2.30) of Ref.~\onlinecite{Goldobin:PhD-Thesis}.

\bibitem{GrE:Stability}
  N.~Gr{\o}nbech-Jensen, D.~Cai and M.R.~Samuelsen,
  Phys. Rev. B {\bf 48}, 16160 (1993).

\bibitem{StkArrDispRel}
  A.~V.~Ustinov, M.~Cirillo, and B.~A.~Malomed,
  Phys. Rev.~B {\bf 47}, 8357 (1993);\\
  A.~E.~Duwel, E.~Tr{\'i}as, T.~P.~Orlando, H.~S.~J. van der Zant,
  S.~Watanabe, and S.~H.~Strogatz,
  J.~Appl. Phys. {\bf 79}, 7864 (1996).

\bibitem{image}
  Yu.~S.~Kivshar, B.~A.~Malomed,
  Phys. Rev.~B {\bf 37}, 9325 (1988);
  Yu.~S.~Kivshar, B.~A.~Malomed,
  Rev. Mod. Phys. {\bf 61}, 763 (1989).

\bibitem{Abramowitz&Stegun}
  M.~Abramowitz and I.~Stegun (Editors),
  ``Handbook of mathematical functions'',
  Dover Publications Inc., New York 14, N.Y.

\bibitem{Vernik:Ic(H)-Ann}
  I.~V.~Vernik, S.~Keil, N.~Thyssen,
  T.~Doderer, A.~V.~Ustinov, and H.~Kohlstedt,
  R.~P.~Huebener,
  J.~Appl. Phys. {\bf 81}, 1335 (1997).

\bibitem{Ust-stack:93}
  A.~V.~Ustinov, H.~Kohlstedt, M.~Cirillo,
  N.~F.~Pedersen, G.~Hallmanns, and C.~Heiden,
  Phys. Rev.~B {\bf 48}, 10614 (1993).

\bibitem{Kleiner2D}
 R.~Kleiner,
 Phys. Rev. B {\bf 50}, 6919 (1994).

\bibitem{HexFisherCherry}
  G.~Hechtfischer, R.~Kleiner, A.~V.~Ustinov and P.~M\"uller,
  Phys. Rev. Lett. {\bf 79}, 1365 (1997).

\bibitem{Krasnov:10-Analytics}
  V.~M.~Krasnov and D.~Winkler,
  Phys. Rev. B {\bf 56}, 9106 (1997).

\bibitem{Goldobin:PhD-Thesis}
  E.~Goldobin, Ph.D. Thesis, in Russian. available online:\\
  http://www.geocities.com/Broadway/2442/

\end{thebibliography}
}
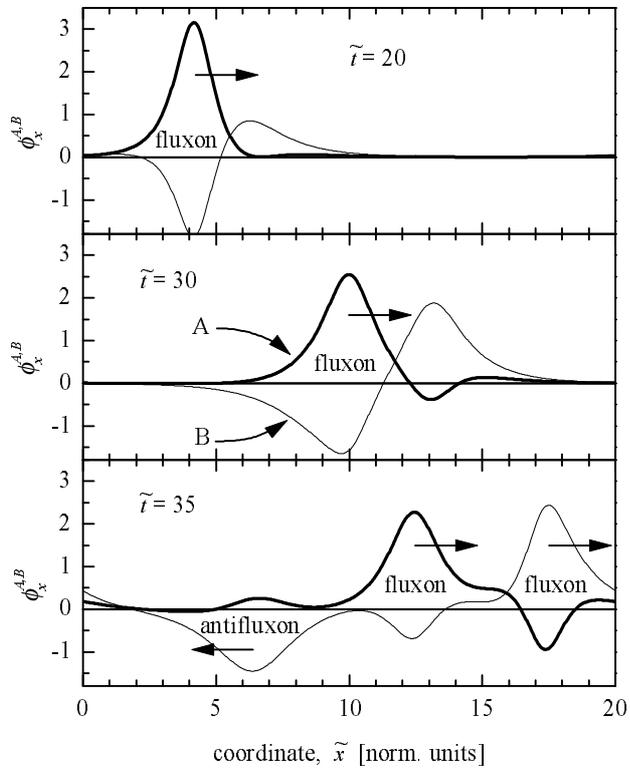
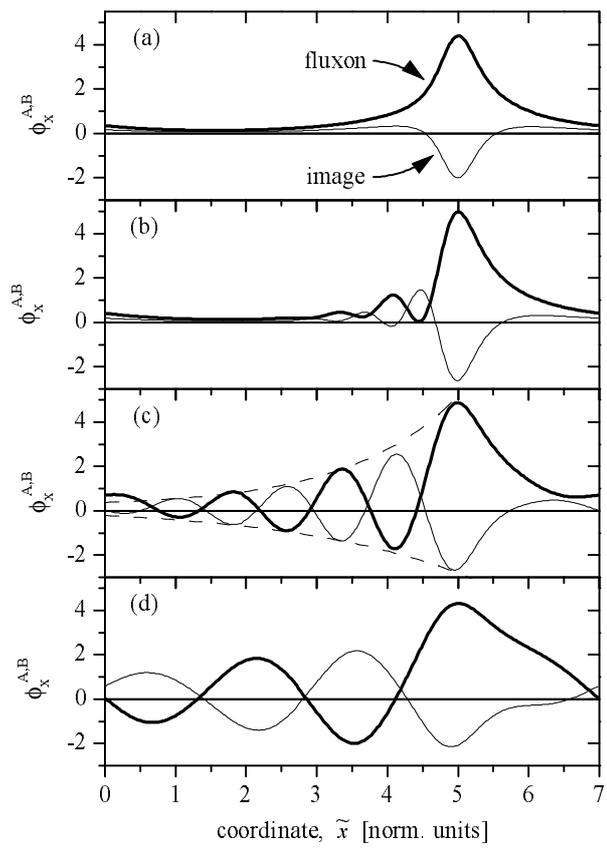
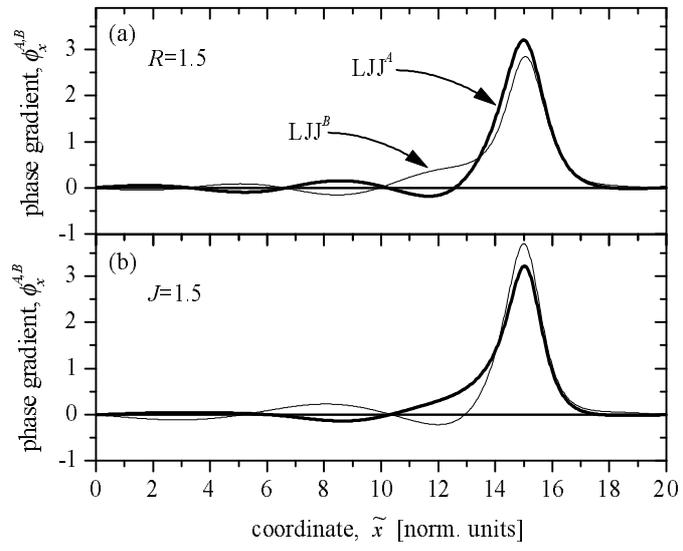
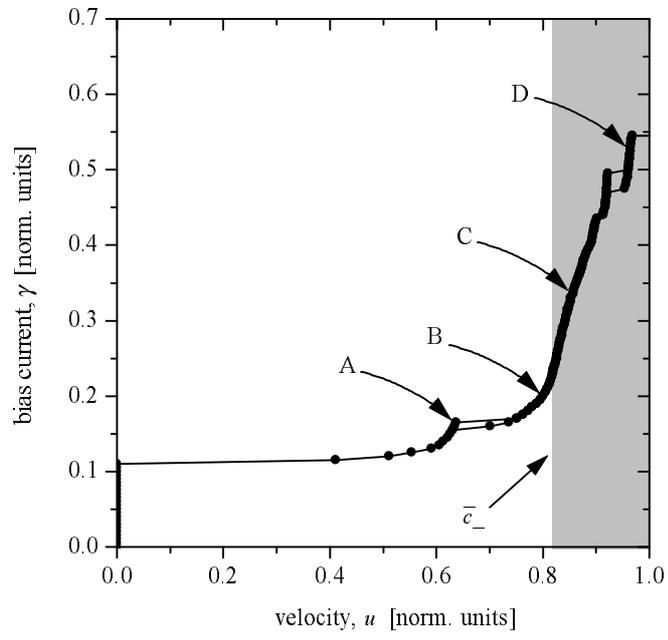
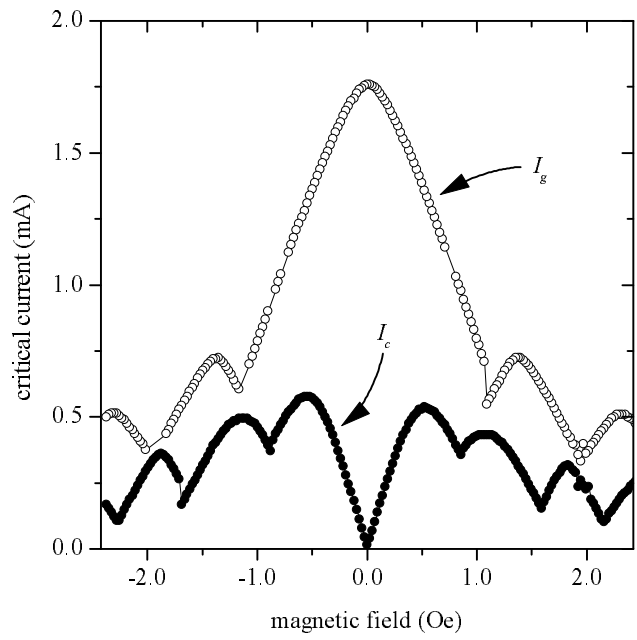
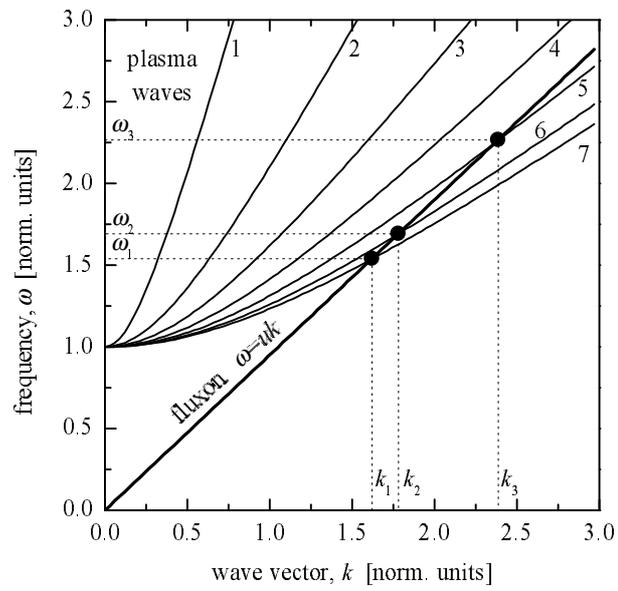
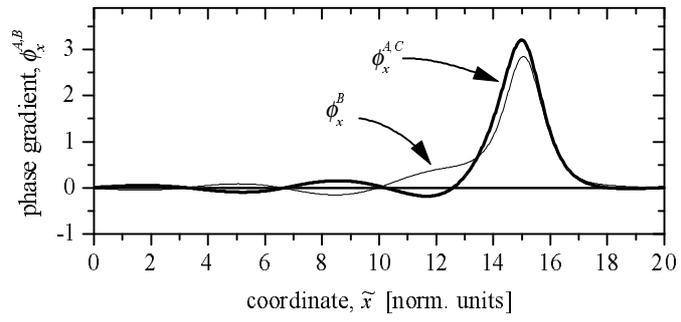
\end{document}